\documentclass[reprint, prc, amsmath,amssymb,
 aps,aps,floatfix,nofootinbib,nobibnotes]{revtex4-2}
\usepackage{epsfig}
\usepackage{color}
\usepackage{amssymb}
\usepackage{amsmath}
\usepackage[ngerman,english]{babel}
\usepackage{multirow}
\usepackage{graphicx}
\usepackage{subfigure}

\usepackage{dcolumn}
\usepackage{bm}
\usepackage{hyperref}

\newcommand{\GR}[1]{\textcolor{cyan}{#1}}

\begin{document} 

\title{
Deuteron yields from LHC: Continuum correlations and in-medium effects}

\author{Benjamin D\"{o}nigus}
\email[Corresponding author:]{benjamin.doenigus@cern.ch}
\affiliation{Institut f\"{u}r Kernphysik, Goethe Universit\"{a}t Frankfurt, Max-von-Laue-Str. 1, 60438 Frankfurt am Main, Germany}

\author{Gerd R\"{o}pke}
\email{gerd.roepke@uni-rostock.de}
\affiliation{%
 ExtreMe Matter Institute EMMI, GSI Helmholtzzentrum f\"{u}r Schwerionenforschung, Planckstrasse 1, 64291 Darmstadt, Germany \\
   and \\
  Institut f\"{u}r Physik,  Universit\"{a}t Rostock, 18051 Rostock, Germany 
}%

\author{David Blaschke}
\email{david.blaschke@gmail.com}
\affiliation{Institute of Theoretical Physics, University of Wroclaw, \\
Max Born place 9, 50-204 Wroclaw, Poland}

\date{\today}

\begin{abstract}
To explain the production of light nuclei in heavy-ion collisions at extreme energies, we focus on the deuteron case. A Gibbs ensemble at chemical freeze-out 
is a prerequisite to investigate the non-equilibrium evolution of the expanding fireball. Quantum statistical approaches allow to describe correlations including bound state formation in the strongly interacting and hot system. 
We consider the virial approach to evaluate proton-neutron correlations. In generalization of the treatment of protons in pionic matter (pion-proton puzzle), the influence of the pion environment on deuteron-like correlations is evaluated using data for the pion-deuteron scattering phase shifts. 
Calculated yields for deuteron production are compared with the ones observed at the LHC.
\end{abstract}

\pacs{21.80.+a, 03.75.Ss, 05.30.âd}

\maketitle

\section{Introduction}

Heavy-ion collisions (HICs) at the Large Hadron Collider (LHC) at CERN produce matter and antimatter with extreme concentration of energy, in a so-called fireball, at mid rapidity shortly after collision. Properties (e.g. composition, momentum distribution of components, etc.) of this extreme matter are reconstructed from measured yields, transverse-momentum spectra and correlations~\cite{ALICE:2019nbs,ALICE:2017jsh,ALICE:2018mdl,ALICE:2017jmf,ALICE:2017nuf,ALICE:2015tra,ALICE:2015oer,ALICE:2015xmh,ALICE:2015wav,ALICE:2014hpa,ALICE:2013xmt,ALICE:2013cdo,ALICE:2013mez,ALICE:2012ovd,Andronic:2017pug,Braun-Munzinger:2018hat,Donigus:2020fon,ALICE:2022amd}. 
The production yields of composite particles, i.e. light (anti-)(hyper-)nuclei, in Pb-Pb collisions are very successfully explained by thermal~\cite{Braun-Munzinger:2003pwq,Andronic:2017pug,Donigus:2020ctf} and coalescence models~\cite{Csernai:1986qf,Mattiello:1995xg,Bleicher:1995dw,Nagle:1996vp,Scheibl:1998tk,Steinheimer:2012tb,Sun:2015jta,Sun:2015ulc,Botvina:2017yqz,Bellini:2018epz,Sombun:2018yqh,Zhao:2018lyf,Sun:2018mqq,Kachelriess:2019taq,Donigus:2020fon,Sun:2020uoj,Hillmann:2021zgj,Glassel:2021rod,Zhao:2021dka}. 
There, thermodynamic equilibrium is assumed at freeze-out, and this determines a primordial distribution of the components of hot and dense matter. A simple statistical-thermal model, the hadron resonance gas (HRG) model~\cite{Braun-Munzinger:1994zkz,Braun-Munzinger:2003pwq,Cleymans:2011pe,Botvina:2008su,Buyukcizmeci:2012it,Floris2014103,Vovchenko:2018fiy,Donigus:2020ctf} has been used to describe the general features of the particle yields (including nuclei) from the fireball 
produced in central HIC, but cannot reproduce some details. 
Recent experiments deliver data with high precision and small statistical fluctuations so that more details of the HIC process are visible.

A first objection with respect to the use of the thermal model is that a HIC is a non-equilibrium process. Also if we assume local thermodynamic equilibrium, the parameters  
density  $n_C$ of conserved quantum number $C$, temperature $T$, and mean velocity ${\bf v}$ of matter are depending on position and time. A hydrodynamical model may provide us with a
profile $n_C({\bf r},t),{\bf v}({\bf r},t),T({\bf r},t)$ evolving in space and time. In this semi-empirical approach, at a certain freeze-out time, the distribution functions remain fixed up to observation of particles by detectors. 

Alternatively, kinetic equations have been worked out, for instance quantum molecular dynamics (see for instance~\cite{Steinheimer:2017vju,Reichert:2021ljd,TMEP:2022xjg}), to describe the time evolution of the fireball. 
There, the inclusion of quantum correlations can only be done semi-empirically after a quasiparticle approach has been performed, 
for instance using a coalescence model~\cite{Sombun:2018yqh,Kittiratpattana:2020daw,Kittiratpattana:2021tpz,Kittiratpattana:2022knq,Kireyeu:2018zdz,LeFevre:2017ygd,Aichelin:2019tnk,LeFevre:2019wuj,Kireyeu:2022qmv}~\footnote{It is worth to mention that the data from ALICE gives hint of a centrality dependence, that is not expected from the thermal model discussed in this article. This is for instance visible from particle yield ratios, e.g. the d/p ratio, which shows a decrease of the central value with increasing centrality (albeit that the experimental uncertainties are still rather large there, such that the trend could also be flat). Nevertheless, this decrease is expected and can be well explained in transport models such as UrQMD~\cite{Bass:1998ca,Bleicher:1999xi,Petersen:2008dd} or SMASH~\cite{Weil:2016zrk,Staudenmaier:2021lrg} by annihilation processes that become more abundant in central events~\cite{Stock:2018xaj}.}

It should be mentioned that recently, a nearly forgotten idea was re-introduced for the description of cluster production~\cite{Dorso:1992ch,Puri:1998te}. Instead of performing an afterburner-type coalescence after the collisions ceased (which is typically after an extremely large time, e.g. 50 fm/$c$) in the transport model, the dynamics of the nucleons is followed throughout the kinetic description and as soon as two or more nucleons are only a certain distance apart from each other and don't suffer any further scattering, the object ''formed'' is treated as a bound cluster~\cite{LeFevre:2017ygd,LeFevre:2019wuj,Aichelin:2019tnk,Kireyeu:2022qmv,Bratkovskaya:2022vqi}. These approaches are very successful in describing the transverse momentum spectra, flow observables as $v_1$ and $v_2$ but also the integrated production yields. Whereas the present work can only cope with the latter, since the prediction of transverse momentum spectra in a statistical-thermal model always requires a set of additional assumptions, e.g. a blast-wave like radial flow~\cite{Schnedermann:1993ws,Danielewicz:1992mi}. In this work we will not address the transverse momentum spectra but only the yields reflecting the chemical composition.

A fundamental approach to the time evolution is given by the non-equilibrium statistical operator $\rho(t)$. 
According to the method of non-equilibrium statistical operator~\cite{Zubarev} it can be constructed from a relevant distribution $\rho_{\rm rel}(t)$,
\begin{equation}
\label{rhoZ}
\rho(t)=\lim_{\epsilon \to 0} \epsilon \int_{-\infty}^{ t} d t' e^{-\epsilon ( t- t')}
e^{-i H ( t- t')}\rho_{\rm rel}( t')e^{i H ( t- t')}
\end{equation}
beeing a solution of the von Neumann equation. 
The choice of $\rho_{\rm rel}(t)$ depends on the information necessary to describe the non-equilibrium state. An application of the Zubarev approach to the problem of pion production in heavy-ion collision experiments has recently been given in~\cite{Blaschke:2020afk}. Up to freeze-out, we
consider the hydrodynamic description as relevant distribution $\rho_{\rm rel}(t)$.
With decreasing density, the relaxation to local thermodynamic equilibrium becomes less efficient, and the freeze-out time is determined by the condition 
that interaction processes are no longer able to sustain the local thermodynamic equilibrium. 
With respect to reactive collisions which change the particle numbers of the components, this is denoted as chemical freeze-out. 
Elastic collisions remain possible at lower densities, they define the kinematic freeze-out. At freeze-out, one has to change the 
description of the non-equilibrium process because instead of the thermodynamic parameters more information is necessary 
to construct the relevant distribution, i.e. the concentration of the different components after chemical freeze out, 
or the single-particle distribution function after kinematic freeze-out. 
Then, the time evolution of the non-equilibrium system is described by reaction kinetics or kinetic equations as performed within transport model calculations. 

Note that this transition from hydrodynamic to kinetic theory is not connected with a change of the physical process but only a question 
of the accuracy in the description if approximations are performed. At freeze-out the deviations from the local equilibrium or the relevant distribution become significant 
so that they have to be treated as new degrees of freedom. In practice, this change of the relevant description is performed 
assuming local thermodynamic equilibrium until freeze-out, and after that the reactions are described by the feed-down from excited states to the particles observed in the experiment. For a systematic non-equilibrium approach to nuclear processes see also \cite{Ropke:2020hbm}. 

However, a systematic treatment of the spectral function is possible at present only in equilibrium. The fact that correlations are important to describe the different dynamical properties of clusters was discussed for instance already in Ref.~\cite{Gossiaux:1994jq}. In case of transport models the quantum mechanical correlations are often lost or in some models not even existent. A first attempt was done in~\cite{Danielewicz:1991dh} and recently also the AMD model tries to include these correlations~\cite{Ono:2019jxm,Ono:2020zsx}. These models are successful mainly at lower energies (few GeV collision energy) and not at the LHC which is the scope of this article.

In any case, an accurate description of the state of hot and dense matter in local thermodynamic equilibrium, i.e. $\rho_{\rm rel}(t)$, is mandatory as prerequisite 
to formulate the non-equilibrium evolution of the system.
Note that, in principle, $\rho_{\rm rel}(t)$ has no influence on the final result if the limit $\epsilon \to 0$ is exactly performed. 
Missing correlations are produced dynamically solving the time evolution operator.
However, in all calculations, approximations are indispensable, and a better choice of  $\rho_{\rm rel}(t)$ gives good results also in lowest approximation.
Occupation numbers of single-particle states may be used to construct $\rho_{\rm rel}(t)$ to describe the evolution after freeze-out.
This leads to kinetic equations, but has problems to incorporate correlation effects. 
Therefore kinetic theory is not appropriate to describe the state of the system before freeze-out. 

A second objection refers to the use of a simple statistical model, the hadron resonance gas, 
describing hot and dense matter in thermodynamic equilibrium as a mixture of non-interacting (with exception of reactive collisions) constituents.
A better description should consider the effects of hadron-hadron interaction, and possible approaches are virial expansions known from nuclear physics~\cite{Ropke:1982ino,Ropke:1982vzx} which are related to the Beth-Uhlenbeck approach \cite{Beth:1937zz} as also shown in~\cite{Horowitz:2005nd,Vovchenko:2017drx} neglecting in-medium corrections. 
The relativistic generalization is given by the S-matrix approach \cite{Dashen:1969ep}.
A particular problem is the treatment of correlations in the continuum which demands a systematic, quantum statistical approach. This approach has been successfully applied to solve the proton puzzle \cite{Andronic:2018qqt}. It was also applied to the strangeness enhancement observed in small collision systems at the LHC \cite{Cleymans:2020fsc} using a coupled-channel approach for the involved phase shifts~\cite{Lo:2017lym,Lo:2020phg}. We are interested in the application to further (composite) hadronic states where the yields are measured, in particular deuterons and antideuterons.

The experimental data we are interested in, are mainly central Pb-Pb collisions at LHC conditions~\cite{ALICE:2013mez,ALICE:2013cdo,ALICE:2013xmt,ALICE:2014jbq,ALICE:2015wav,ALICE:2015oer,ALICE:2017ban,ALICE:2017jmf}. 
For instance, at collision energy $\sqrt{s_{NN}}=2.76$ TeV, a fireball is produced at midrapidity. At chemical freeze-out,  it is characterized  by the grand-canonical distribution $\rho_{\rm rel}(t)$ with baryonic chemical potential  $\mu_B \approx 0 $, radius about 10.5 fm, corresponding to a volume of about $V_{\rm fr} \approx 4200$ fm$^3$, 
and a temperature of about $T_{\rm fr}=156$ MeV. The number of measured charged pions per rapidity unit $dN_{\pi}/dy$ is about 700, for all three species leading to roughly 2100, what corresponds to a pion density of about 0.34 fm$^{-3}$.
The rapidity density $dN/dy$ for a particle species $i$ at mid-rapidity can be written as (see for instance~\cite{Schnedermann:1993ws})
\begin{equation}
\label{dndy}
dN_i/dy|_{y=0} = \frac{g V_{\rm fr}}{(2\pi)^2}\int dp_T p_T \sqrt{p_T^2 + m_i^2}e^{-\sqrt{p_T^2 + m_i^2} / T_{\rm fr}},
\end{equation}
where $g$ is the spin-isospin degeneracy factor. 
The integration is over the transverse momentum  $p_T$ of the particle $i$ with the mass $m_i$. Eq.~(\ref{dndy}) already assumes that the baryo-chemical potential $\mu_B$ is zero~\cite{Andronic:2017pug}. 
The fact that $\mu_B=0$ requires that particles and antiparticles are produced with equal weight. It also means that the net-baryon density is zero.
The numbers of protons/antiprotons and neutrons/antineutrons within this fireball are about 15 per species primordially, and close to 40 after taking into account the feed-down, each so that the baryon density 
is about $3 \times 10^{-2}$  fm$^{-3}$, i.e. very low. 
In a first approach the hadronic resonance gas has proven to be surprisingly efficient to describe the observed yields. Nevertheless, the new experiments give yields of different particles with high precision so that improvements of this simple model, in particular the effect of interactions, have to be taken into account to explain the data.
 
\begin{figure}[htb]
    \begin{center}
    \includegraphics[width = 0.48\textwidth]{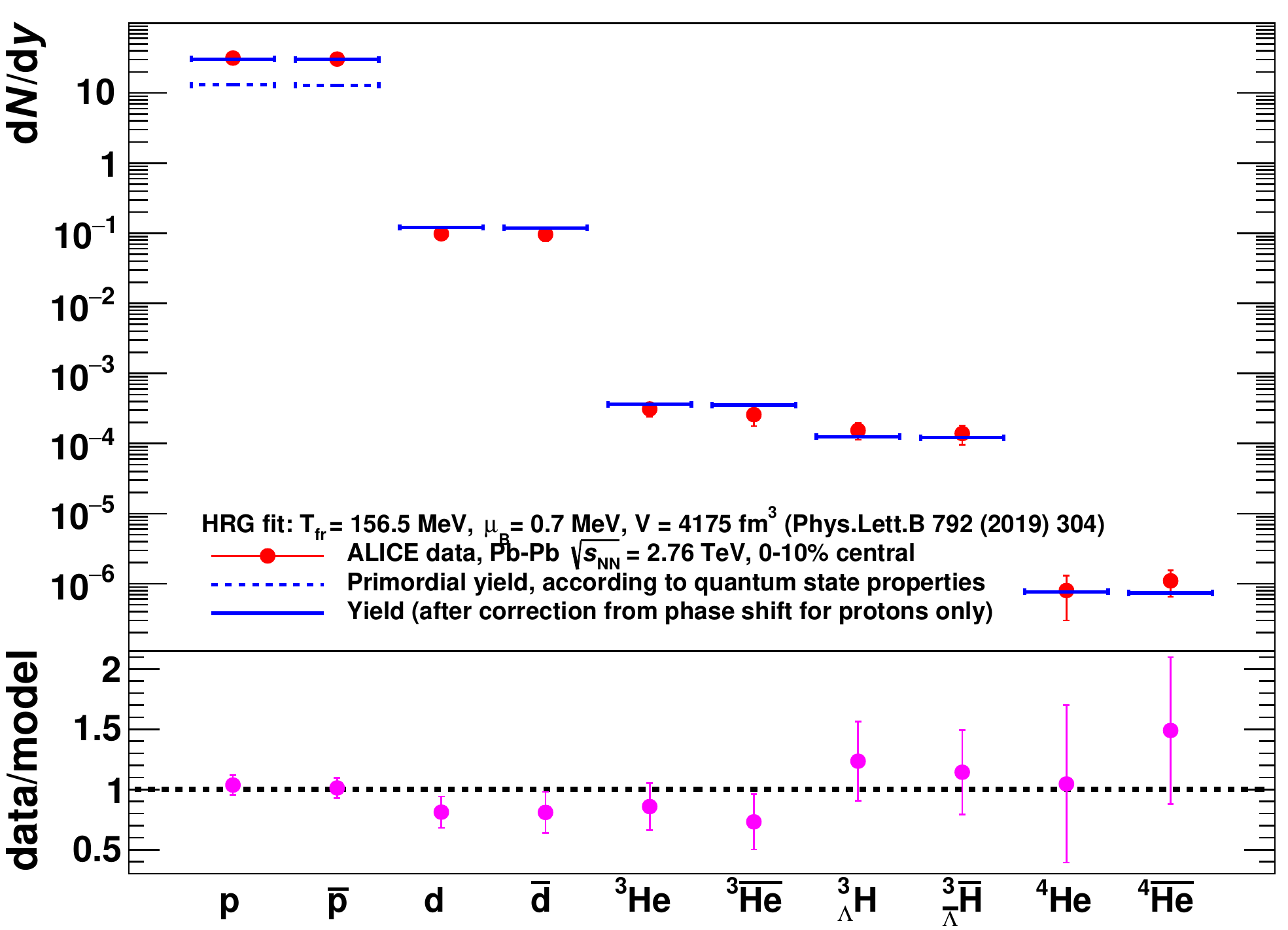}
    \end{center}
    \label{fig:pbm}
    \caption{Result of the thermal model fit to ALICE data ($\sqrt{\it{s}_{NN}}=2.76$~TeV, 0-10\% central) on particles per rapidity unit $dN/dy$ for non-strange (anti-)baryons and (anti-)(hyper-) nuclei using the S-matrix approach discussed below (for details of this fit see~\cite{Andronic:2018qqt}). In detail, the figure shows in the upper panel the data (red) compared with the fit result, separated in primordial yield (dashed blue line) and final result (solid blue line) after feed-down and S-matrix correction for the protons. In addition, the data-over-model ratio is shown in the lower panel (in magenta), using the uncertainties of the data. One clearly sees that the protons and antiprotons are well described after the S-matrix correction. Whereas light nuclei are showing the same tension the protons showed before the correction (see \cite{Andronic:2017pug}). Nevertheless, the uncertainties of light nuclei are still large enough to cope with the 1$\sigma$ deviation, corresponding to 20-30\% difference between fit and data.}
\end{figure}
 
 The aim of this work is to investigate the modification of the production yields of deuterons and antideuterons if a quantum statistical approach is used instead of a simple HRG. Any further comparison is not in the scope of this publication, since it would involve additional assumptions. For instance, to describe the transverse momentum spectra of protons and deuterons we would need at least the mean flow velocity $\langle\beta\rangle$ and the surface parameter $n$ for a blast-wave like model~\cite{Schnedermann:1993ws} which nicely incorporates the features of radial flow, that is needed to be taken into account when (central) heavy-ion collisions are investigated.
 Experimental data for the corresponding production yields from the ALICE Collaboration at LHC are given in \cite{ALICE:2015wav,ALICE:2013mez,ALICE:2012ovd,Andronic:2018qqt}. Figure~\ref{fig:pbm} shows the experimental data on particles per rapidity unit $dN/dy$ for for non-strange (anti-)baryons and \mbox{(anti-)}(hyper-)nuclei measured by the ALICE Collaboration in Pb-Pb collisions at $\sqrt{\it{s}_{NN}}=2.76$~TeV in a centrality interval of 0-10\%, compared to the thermal model fit described in~\cite{Andronic:2018qqt}. The fit using a 
 hadronic resonance gas overestimates the measured deuteron and antideuteron yields by about 20\%. 
 
 We consider improvements taking into account continuum correlations and interaction with the hot and dense surrounding matter.
According to the composition given above, we expect that the main effects are caused by the pions, i.e. we have hadronic clusters embedded in 
hot and dense pionic matter. Consequently the main contribution to in-medium modification of hadrons is due to the interaction with pions,
and the knowledge of corresponding scattering properties like phase shifts is necessary to calculate the in-medium effects on the composition. In our work, the partial density of deuteron-like correlations is investigated.

We start with the treatment of the interacting many-particle system and the introduction of partial densities in Sec. \ref{Partial}.
We give in Sec. \ref{sec:proton} a short reference to the modification of the proton yield because of the interaction with surrounding pions as performed in Ref.~\cite{Andronic:2018qqt}. In Sec. \ref{Continuum} we discuss the formation of deuterons and the influence of continuum correlations, 
and in Sec. \ref{Phaseshifts} we calculate the 
modification of the deuteron yield owing to the interaction with surrounding pions. Conclusions are drawn in Sec. \ref{Conclusions}.

\section{Partial intrinsic partition functions}
\label{Partial}
Hot and  dense matter can't be described as ideal quantum gas. Because the components are interacting, correlations are formed. Examples are bound nuclei, in the ground state and excited states, but also resonances and continuum correlations. 
In this section, we demonstrate how the treatment of interactions for nuclear matter consisting of protons and neutrons allows to explain the formation of deuterons. We consider clusters of $A$ nucleons which are characterized by total momentum $P$ and further quantum numbers like isospin, 
here the proton number $Z$, and  angular momentum $J$.
Starting from a general quantum statistical approach, we decompose the total density in partial contributions from the different channels characterized by $A, Z, J$~\cite{Ropke:2014fia},
\begin{equation}
\label{ntotal}
n_\tau^{\rm total}(T,\mu_n,\mu_p)=\sum_{A,Z,J} A_\tau n_{A,Z,J} ^{\rm part}(T,\mu_n,\mu_p),
\end{equation}
$\tau$ denotes neutron ($n$) or proton ($p$), and $A_\tau$ the neutron number  ($A-Z$) or proton number ($Z$) of the cluster.
The partial densities of the different channels are given by (non-degenerate case)
\begin{eqnarray}
\label{npart}
&&n_{A,Z,J} ^{\rm part}(T,\mu_n,\mu_p)=e^{((A-Z)\mu_n+Z \mu_p)/T}\nonumber\\
&& \times \int \frac{d^3P}{(2 \pi)^3} e^{- P^2/(2Am_NT)}
z_{A,Z,J} ^{\rm part}(P;T,\mu_n,\mu_p).
\end{eqnarray}
$m_N$ is the nucleon mass and $Am_N$ is the mass of the cluster $\{A, Z\}$. Degeneracy effects are not relevant at conditions considered here, the  Boltzmann probability distribution of the  cluster states is given by the energy. After separation of the kinetic energy of the center-of-mass motion, the intrinsic partition function $z_{A,Z,J} ^{\rm part}$ contains all intrinsic excitations of the cluster.

We give the  contributions of clusters with lowest mass number $A$. For $A=1$ we have the contributions of free neutrons and protons, $J$ is replaced by the spin direction, 
there are no intrinsic excitations within the hadronic phase at temperatures small compared to the energy of resonances so that the intrinsic partition function  is $z^{\rm part}_{A=1}=1$.
Within an advanced approach, the nucleon mass $m_N$ should be replaced by the quasiparticle mass which contains the effect of a mean field. Well-known are, e.g., the relativistic mean-field approximations obtained from  model Lagrangians which describes the coupling of the nucleons to  mesonic fields, see for instance~\cite{Typel:2009sy}.

For $A=2$ we have isospin triplet ($nn, np, pp$) channels as well as the isospin singlet ($np$) channel where the deuteron ($d$) is found as bound state. Therefore, this channel is of particular interest in our present work.
For the corresponding $z_{np}^{\rm part}$, the  sum over the intrinsic excitations includes the continuum of scattering states. We replace the sum over the continuum states by the integral over the laboratory energy $E^{\rm lab}$ of the colliding nucleon. The Beth-Uhlenbeck formula~\cite{Beth:1937zz,Ropke:1982vzx,Ropke:1982ino} is obtained which reads for the isoscalar channel where the deuteron is found
\begin{eqnarray}
\label{BU}
&&z_{np} ^{\rm part}(P;T,\mu_n,\mu_p) =
3 e^{B_d/T}\nonumber \\ &&
+\frac{1}{\pi}\int dE^{\rm lab} e^{-E^{\rm lab}/2T}\frac{d }{d E^{\rm lab}}\delta_{np}^{\rm tot}(E^{\rm lab}).
\end{eqnarray}
 The deuteron binding energy $B_d=2.225$ MeV (degeneracy factor 3 according to spin 1) and the scattering phase shifts $\delta_{np}^{\rm tot}(E^{\rm lab})$ are known from experiments, see also~\cite{Horowitz:2005nd}.
A generalized form of the Beth-Uhlenbeck formula~\cite{Schmidt:1990oyr} which accounts for in-medium effects is presentd in the Appendix~\ref{lab:appendix}.
The treatment of the second virial coefficient is given below in Sec. \ref{Continuum} where the deuteron formation is considered.

\section{Protons in pion matter}
\label{sec:proton}

The Beth-Uhlenbeck approach can be generalized to other components of the many-body system. In this section, we are interested in the interaction of protons with pions as the main component in the fireball. The approach which has been worked out for interacting baryon systems will be applied here for a system mainly consisting of pions so that the interaction of nucleons with pions is the main effect. We present this issue here to compare with the work~\cite{Andronic:2018qqt}.

We briefly repeat  the treatment of nucleons in pion matter. The hadron resonance gas would consider a mixture of pions, nucleons, $\Delta$ and other particles as listed, e.g., in the particle data book \cite{Zyla:2020zbs}. The primary yield ratio of $\Delta$ resonances ($m_\Delta=1232$ MeV) to nucleons ($m_N=939$ MeV) is for $T_{\rm fr}=156$ MeV:
\begin{equation}
\label{Delhg}
\frac{Y^{\rm prim}_\Delta}{Y^{\rm prim}_n+Y^{\rm prim}_p}=\left( \frac{m_\Delta}{m_N}\right)^{3/2} \frac{16}{4} e^{-(m_\Delta-m_N)/T_{\rm fr}}=0.919.
\end{equation}
These $\Delta$ resonances which are present in thermodynamic equilibrium at freeze-out, will disintegrate during the expansion of the hot and dense matter.
Because baryon number is conserved, their decays feed into the nucleon and pion channels. Therefore, the final yields of the nucleon $\tau$ is increased compared to the primary yield by a factor of 1.919. In particular we expect within the hadron resonance gas model 
$Y_p^{\rm res.g. \Delta}=1.919 \,Y_p^{\rm prim}$ only taking into account the contribution of the $\Delta$ resonances.

\begin{figure}
\begin{center}
 \includegraphics[width = 0.46\textwidth]{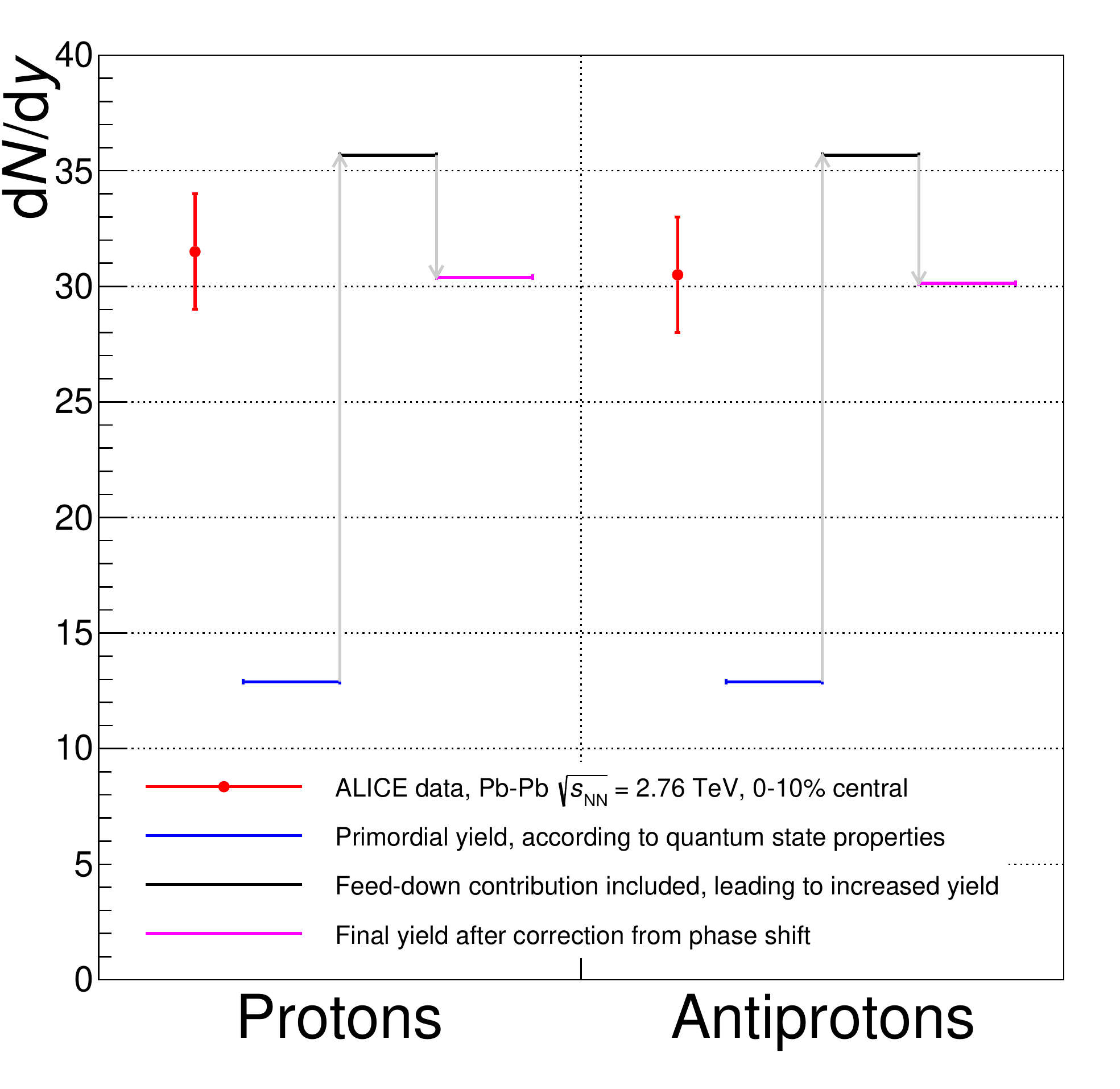}
    \end{center}
     \caption{Comparison between experimental data from Pb-Pb collisions at $\sqrt{s_\mathrm{NN}} = 2.76$~TeV for protons and antiprotons with values of model calculations. The red points indicate the production yield d$N$/d$y$ and the vertical lines attached to them the quadratic sum of the statistical and systematic uncertainties. The blue horizontal lines indicate the primordial yields, the black lines the yields corrected for feed-down contributions and in magenta the yields after correcting the resonance contribution by the phase-shift analysis through the virial approach.}
\label{fig:2}
\end{figure}

Additional resonances, in particular $N^*(1520)$, will also contribute. The HRG gives a factor 0.1498 in addition to the $\Delta$ resonances. 
The sum over all resonances given by the PDG~\cite{Zyla:2020zbs} increases the final proton yield by the factor 2.743. This is shown in Fig. \ref{fig:2} 
where the primordial proton yield according to the temperature $T_{\rm fr}=156$ MeV 
\begin{equation}
\label{proton_yield}
\frac{dN_p^{\rm HRG}}{dy} = 12.894, 
\end{equation}
is increased within the HRG model to 35.668 (including the feed-down from decays of primordial $\Delta$(1232) with multiplicity 10.839 which is a substantial contribution).

However, this statistical hadronization model predicts about 20\% more protons and antiprotons ($dN_{\rm p}/dy$ = 36 instead of 30) than measured by the ALICE Collaboration in central Pb-Pb collisions at the LHC. 
This constitutes the much debated proton-yield anomaly (also called proton ''puzzle'')  in heavy-ion collisions at the LHC \cite{Andronic:2018qqt}. A possible approach to resolve  this anomaly was to improve the non-interacting HRG model by using exact expressions for the second virial coefficient given by Dashen, Ma, and Bernstein \cite{Dashen:1969ep} for the second virial coefficient, containing the pion-nucleon phase shifts. 

We shortly repeat the calculation of the virial expansion of the density where the second virial coefficient is calculated within the Beth-Uhlenbeck approach \cite{Beth:1937zz}.
In addition to the single-particle contributions describing the ideal gas of neutron and proton quasiparticles,   
the pion-nucleon channels are considered. They contain the $\Delta$ resonance seen in the $P_{33}$ channel,
as well as further excited states. 
The phase shifts $\delta_l$ for different channels are well known, fitted analytical expressions are available \cite{Rowe:1978fb}. 
The intrinsic partition function $z_{A,Z,J} ^{\rm part}$ follows in the Boltzmann limit as~\cite{Weinhold:1997ig}
\begin{eqnarray}
\label{nucleon}
z_{1,Z,l} ^{\rm part}=\frac{1}{\pi T}
 \int {dE} e^{-{E}/T} \delta_l(E).
\end{eqnarray}
In the Appendix, a more general form of the Beth-Uhlenbeck formula (Eq.~(\ref{gBU})) is given which accounts for the introduction of quasiparticles. We use the free phase shifts
to evaluate the expression and obtain the second virial coefficient. In-medium modifications of the scattering phase shifts, as performed in~\cite{Schmidt:1990oyr}, are not considered in this work. 

The energy-dependent phase shifts of pion-nucleon scattering below 400 MeV have been fitted in \cite{Rowe:1978fb} as
\begin{equation}
\tan \delta_l=q^{2l+1}\left[b+c q^2+d q^4+\frac{x \Gamma_0 \omega_0 q_0^{-(2l+1)}}{\omega_0^2-\omega^2}\right]\,.
\end{equation}
(This fit is extended to  energies larger than 400 MeV but gives only a contribution of few percent so that the error will be small. This is acceptable for this study, since we mainly want to show that the Beth-Uhlenbeck approach used here gives similar results as Dashen-Ma-Bernstein approach in Ref.~\cite{Andronic:2018qqt}).
For an exploratory calculation, we consider only the $P_{33}$ channel which gives the dominant contribution.
According to~\cite{Rowe:1978fb} we use for the $P_{33}$ phase shift $(l=1)$ the values $b=0.114/m_\pi^3, \,\,\,c=-0.0154/m_\pi^5,\,\,\, d=0.00072/m_\pi^7, \,\,\,x=0.99,\,\,\, \Gamma_0=116, \,\,\, \omega_0 =1232,\,\,\, q_0=228$ (units MeV, MeV/c) and with $m_\pi$ being the pion mass.
For the freeze-out temperature $T_{\rm fr}$ = 156 MeV, we obtain using Eq.~(\ref{nucleon}) for the proton using the pion-proton channel:
\begin{equation}
\label{sol_nucleon}
z^{\rm part}_{1,1,1}=\frac{1}{\pi T} \int_0^{\infty} dE e^{-E/T_{\rm fr}} \delta_1(E) = 0.388. 
\end{equation}
The account of the sin-term appearing in the generalized Beth-Uhlenbeck formula Eq.~(\ref{gBU}) reduces this to 0.372 (about 5\%) what indicates that this generalization gives here only a small effect. Note that phase shift must be given in radians.

Compared to the HRG result $e^{-(m_\Delta-m_N-m_\pi)/T_{\rm fr}}=0.374$ (note that the continuum edge contains also the pion mass $m_\pi=139.6$ MeV) we have nearly the same result. 
With the multiplicity $4 (m_\Delta/m_N)^{3/2}$ given in Eq.~(\ref{Delhg}), we obtain the amount of  $\Delta$ resonances which decay to nucleons when the 
hot matter is expanding after freeze-out. These feed-down processes contribute to the final proton yield.

However, other channels also contribute and we have to include higher resonances.  
A systematic calculation has been performed in \cite{Andronic:2018qqt}. We mainly wanted to show that our approach using the generalized Beth-Uhlenbeck formula works as well as the Dashen, Ma and Bernstein \cite{Dashen:1969ep} approach.
We will not repeat these calculations here but only present the main concepts to solve the proton puzzle. As a result, the reduction using the scattering phase shifts gives a total contribution of all proton-pion channels of 30.4 which feeds the final proton yield. The analogous calculation for antiprotons give the value 30.1, the small difference is due to the non vanishing, but small value of the baryonic chemical potential. The actual fit of the measured yields in Pb-Pb collisions at 2.76 TeV at the LHC leads to a value of $\mu_\mathrm{B}= (0.7 \pm 3.8)$~MeV~\cite{Andronic:2017pug}.

Figure~\ref{fig:2} shows all involved steps to infer, from the final experimental values for the  yields of protons and antiprotons, the primordial yields characterizing the composition at freeze-out. Feeding from the pion-proton channels, the primordial proton yield is changed to the  final proton yield which is observed in experiments. The  hadron resonance gas model predicts an enhancement of the primordial yield of the nucleons by a factor of 2.74 owing to the feed-down from hadronic resonances which is reduced to 2.204 using the virial approach, what agrees with the value seen from the final yields in the experiment.

We focus here only on the second virial coefficient and in our approach we do not consider in medium modifications of scattering phase-shifts in a dense environment that would lead to higher virial coefficients.

\section{Deuteron channel}
\label{Continuum}

We are interested in the deuteron production also observed in HIC at the LHC~\cite{Andronic:2018qqt}. The number of measured deuterons per rapidity unit is $dN_d/dy=0.098$ and $dN_{\bar d}/dy=0.092$ for antideuterons.
Within the simple approach of nuclear statistical equilibrium, the ratio of deuteron ($s=1$) yield $Y_d$ to proton ($s=1/2$)  yield $Y_p$ is given as 
\begin{equation}
R^{\rm HRG}_{dp}=\frac{Y_d}{Y_p}=\frac{3}{2} \,\frac{ \int d^3p/(2 \pi)^3 e^{-\sqrt{m_d^2+p^2}/T}}{\int d^3p/(2 \pi)^3 e^{-\sqrt{m_p^2+p^2}/T}}
\label{HRG}
\end{equation}
assuming $\mu_B=0$. With $T_{\rm fr}=156$ MeV the value $R^{\rm HRG}_{dp}=0.00908$ follows. Together with the value 12.894 from Eq.~(\ref{proton_yield}) for the proton yield given above, the deuteron yield is 
\begin{equation}
\label{deuteron_yield}
dN^{\rm HRG}_d/dy = 0.1171,    
\end{equation}
see also Fig. \ref{fig:3}. Compared to the measured yields, the HRG model overestimates the deuteron production from HIC.

Several issues can be given which improve this simple statistical approach. As discussed at the beginning, the time evolution of the hot and dense matter produced in HIC collision is described by the statistical operator $\rho(t)$. 
This statistical operator $\rho(t)$ is formulated using the relevant statistical operator (Gibbs distribution) $\rho_{\rm rel}(t')$ 
\begin{equation}
\rho_{\rm rel}=\frac{e^{-\beta (H-\sum_i \mu_i N_i)}}{{\rm Tr}\, e^{-\beta (H-\sum_i \mu_i N_i)}}
\end{equation}
where $H$ is the Hamiltonian of the system, $N_i$ the particle number of conserved components $i$. 
In general, the Lagrange parameters $\beta, \mu_i$, denoting the inverse temperature and the chemical potentials, are depending on  position and time. 
In this context, the HRG Eq.~(\ref{HRG}) appears as a simple approximation where the Hamiltonian $H$ is replaced by an expression where all interactions are neglected, after bound states have been introduced with the corresponding binding energies. An ideal, non-interacting mixture of free nucleons and bound states is considered, with accidental reactions and collisions to sustain partial equilibrium.
We improve this in this work taking interactions into account as well as excited states, including continuum correlations.

A systematic quantum statistical approach introduces the spectral function which contains all correlations in the hot and dense nuclear matter.
Two-particle correlations including the deuteron are obtained from the two-particle propagator, as solution of a Bethe-Salpeter equation.
In contrast to the proton, the deuteron is a composite particle consisting of a neutron and a proton, with a 
binding energy $B_d=2.225$ MeV. There are no excited bound states, but there are correlations in the continuum as described by the $n - p$ scattering phase shifts. As shown from the Beth-Uhlenbeck formula Eq.~(\ref{BU}), the total amount of density in the isoscalar channel, where the deuteron is found, is described by the second virial coefficient $b_{np}(T)=z_{np}^{\rm part}(T)/\sqrt{2}$, 
\begin{equation}
n_{np}^{\rm part}(T,\mu_n,\mu_p)
=\frac{4}{\Lambda^3}e^{(\mu_n+\mu_p)/T} b_{np}(T),
\end{equation}
where $\Lambda=[ 2 \pi /(m_NT)]^{1/2}$ is the thermal wave length of the nucleon, with mass $m_N$.
 An expression for the second virial coefficient \GR{(\ref{gBU})}  including in-medium effects is obtained from the generalized Beth-Uhlenbeck formula~\cite{Schmidt:1990oyr} which can be used also in the high-density region, see Appendix~\ref{lab:appendix}.
 
 Nevertheless, there are excited states of ${}^{4}$He, ${}^{5}$He and ${}^{5}$Li that are unstable particles (strong decays) and enhance the deuteron yield artificially~\cite{Vovchenko:2020dmv}. These increase the (thermal) production yield by about 0.0225\% as shown in Fig.~\ref{fig:3}. The feeddown is such low since the penalty factor $F$ to produce one of the nuclei is about 330 per additional baryon~\cite{Braun-Munzinger:2018hat}. For the higher mass nuclei mentioned above this means a suppression by $F^{2}$ or even $F^{3}$, i.e. nearly negligible.
 
 We consider the region of low baryon density where in-medium effects can be neglected.
 According to Beth and Uhlenbeck \cite{Beth:1937zz}, 
 the second virial coefficient can be expressed in terms of the binding energy 
 and scattering phase shifts. After integration by parts of Eq. (\ref{BU}) we find
\begin{equation}
\label{deutBU}
b^{\rm vir}_{np}(T)=\frac{3}{2^{1/2}}\left[e^{B_d/T}-1+\frac{1}{\pi T} \int_0^\infty dE e^{-E/T} \delta^{\rm total}_{np}(E) \right].
\end{equation}
Here, $E$ denotes the energy of relative motion, the energy of the center of mass motion of the two-nucleon system has been integrated over.
Note that we use a nonrelativistic approach to introduce bound state wave functions and scattering phase shifts. Relativistic kinematics may be introduced, and the S-matrix approach can be given, but relativistic generalizations of statistical operator and in-medium Schr{\"o}dinger equations demand much more effort.

Equations~(\ref{BU}), (\ref{deutBU}), as well as the general form (\ref{gBU}) given in the Appendix~\ref{lab:appendix} can be rewritten in a compact form without subdivision of bound and scattering state contributions
when a generalized scattering phase shift $\delta_d^{\rm gen}$ is introduced, see~\cite{Ropke:2014mwa,Bastian:2018wfl}.
We define $\delta_d^{\rm gen}(E) = \pi$ for $-B_d \le E\le 0$ if there exists a bound state with a binding energy $B_d$, and $\delta_{np}^{\rm gen}(E) = \delta_{np}(E) $ for $E \ge 0$. For Eq.~(\ref{BU}) we obtain: 
\begin{eqnarray}
&& z_d^{\rm part} (T,\mu_n,\mu_p) = \frac{3}{\pi T} \int_{-\infty}^{\infty} dE e^{-E/T} \delta_{np}^{\rm gen}(E).
\end{eqnarray}
This expression represents the total amount of correlation, avoiding the (artificial) subdivision into 
a bound part contribution and a scattering part contribution. A corresponding expression can also found for the generalized Beth-Uhlenbeck formula, Eq.~(\ref{gBU}).

Using the measured phase shifts, Horowitz and Schwenk \cite{Horowitz:2005nd} calculated values for $b_{np}(T)$ for $1 \le T \le 20$ MeV. We extend these calculations to the quasiparticle picture, see Appendix~\ref{lab:appendix}, and obtain for the freeze-out temperature $T_{\rm fr}=156$ MeV the value $b_{d}^{\rm qu}(T_{\rm fr})=0.971$. This means that the yield obtained from the HRG calculation is reduced by the factor 
\begin{equation}
\label{vird}
b^{\rm qu}_d(T_{\rm fr}) \frac{2^{1/2}}{3} e^{-B_d/T_{\rm fr}}=0.451,
\end{equation} 
so that the production yield of deuterons 
\begin{equation}
\frac{dN_d^{\rm virial}}{dy} = 0.0529 
\end{equation}
results. This value is shown in Fig. \ref{fig:3}.

We conclude that the deuteron is only weakly bound, and considering the spectral function or the Beth-Uhlenbeck formula Eq.~(\ref{deutBU}), most of the contribution to the correlated density in the  isospin singlet channel, spin 1, is obtained from the continuum. The virial coefficient $b^{\rm vir}_{np}(T)$ contains all correlations in the isospin singlet channel where the deuteron is found as the state with lowest energy. Note that these deuteron-like correlations will not necessarily feed the observed deuteron states so that the calculated value d$N_d^{\rm virial}$/d$y$ represents an upper limit. Comparing with the simple nuclear statistical equilibrium yield $dN^{\rm HRG}_d/dy=0.1171$ from Eq.~(\ref{deuteron_yield}), a significant reduction of the deuteron yield is observed if continuum correlations are taken into account.

Like the production yield of protons calculated in the statistical model, the value $dN^{\rm virial}_d/dy$ is very small (about 1/2) compared to the 
measured yield $dN_d/dy=0.098$. The yield of deuteron-like correlations in high-energy density matter according to the virial expansion shown in Fig. \ref{fig:3} underestimates the observed yield, similar to the primary proton yield shown in Fig. \ref{fig:2}. 

The question whether a weakly bound state such as the deuteron, $B_d=2.225$ MeV can survive in a fireball with temperatures of the order 100 MeV, has been discussed in several publications, see for instance \cite{Mrowczynski:2016xqm,Oliinychenko:2018ugs,Vovchenko:2019aoz,Mrowczynski:2020ugu,Gallmeister:2020fiv}. 
Like snowballs in hell~\cite{Braun-Munzinger:1994zkz,cern_courier}, they argue that deuterons cannot survive in the fireball. According to the coalescence model, it is proposed that
light nuclei are formed due to final-state interactions after the fireball decays. 
This means that chemical freeze-out up to which formation processes of deuterons are possible to occur at a later instant of time,
at different thermodynamic parameters.
However, correlations are present also in  matter with high density of energy at freeze-out, as described by quantum statistical approaches.
The  virial coefficient $b^{\rm qu}_d(T_{\rm fr})=0.971$ Eq.~(\ref{vird}) expresses the amount of correlations for the interacting nucleon system at temperature $T_{\rm fr}$.

The virial equation of state which accounts only for nucleon-nucleon interaction cannot explain the observed deuteron yields from the experiments~\cite{Andronic:2018qqt}. 
Similar to the proton case, interaction with the pion system including the formation of resonances have to taken into account.

\section{The deuteron in pion matter}
\label{Phaseshifts}

Simple statistical models like the hadron resonance gas are improved when the interaction between the constituents is taken into account.
Empirical approaches such as the concept of excluded volume (see for instance \cite{Vovchenko:2017xad,Vovchenko:2017drx,Huovinen:2017ogf,Hempel:2009mc,Vovchenko:2020lju,Bugaev:2020sgz} and references therein) are not well founded.  
Before considering the interaction of the nucleons with pions, we shortly discuss the systematic treatment of the interaction of deuterons with other nucleons. 
A quantum statistical approach has been worked out~\cite{Schmidt:1990oyr}, and self-energy shifts and Pauli blocking effects have been investigated.
The shift of the binding energy owing to Pauli blocking at nucleon density $n_p+n_n= 0.015 $~fm$^{-3}$ and $T_{fr}=156$ MeV has been evaluated in \cite{Ropke:2008qk} to be 0.2 MeV. This gives a reduction of about 1 per mille and can be neglected. Only at baryon densities of the order of the saturation density, $n_B \approx 0.1$ fm$^{-3}$, medium effects become relevant. 

However,  we have a large value for the pion density so that the interaction with the pionic environment has to be considered, similar as done above for the case of protons. 
Before we perform a detailed calculation, we give a rough estimate along the lines of the HRG. 
We focus on the $\Delta$ resonances which are dominant because of the low excitation energy and the large statistical factors (spin: 4, isospin: 4).

The deuteron is a weakly bound state of two nucleons which move almost freely. 
If we apply the impulse approximation, both constituents of the deuteron, at given distribution in momentum space according 
to the bound state wave function, are assumed to interact separately with the pion environment. We observe such a behavior in the pion-nucleon cross section \cite{Zyla:2020zbs}
where the pion-deuteron scattering cross section is nearly the sum of the individual pion-nucleon cross sections. 
In particular, this refers also to the large peak near the position of the $\Delta$ resonances.
Because nucleons in pionic matter are dressed forming resonances, we have also such resonances for the nucleons as constituents of the deuteron. 
If we assume that both nucleons forming the deuteron are dressed by pions so that single-nucleon spectral function has peaks near the hadronic resonances such as $\Delta$, 
according to Eq.~(\ref{Delhg}) an enhancement  factor of $1.919^2= 3.682$ would appear. 
Then, the production yield of deuterons would amount to
\begin{equation}
dN_d^{\rm res.gas}/dy = 0.1947.
\end{equation}
Here we assume that nucleon - $\Delta$ correlations behave similar to the nucleon-nucleon correlations which determine the virial 
expansion.  As in the proton case, the $\Delta$ resonances decay after freeze-out to feed the nucleon yields.

Compared with the experimental yield, this estimate is very large already if no higher resonances are taken into account.
It is obvious that such hadron resonance gas approximation is not realistic because we cannot expect that the proton-neutron interaction
coincides with the corresponding baryon-baryon  interaction including the $\Delta$ resonance to form the same correlations. 
The impulse approximation which neglects the energy dependence of the self-energy  has to be improved.

To find a consistent solution, we should describe the deuteron in pion matter in a systematic way, 
as done above for the proton in pion matter to solve the  proton puzzle.

A first-principle approach to describe the deuteron in pion matter can be given considering the spectral function for the proton-neutron propagator in pion matter. 
As well known, the deuteron appears as a pole of this propagator in ladder approximation, 
solving the Bethe-Salpeter equation.
The interaction with the pion environment is described by a self-energy, 
and a Beth-Uhlenbeck formula can be derived which expresses the density in terms of the deuteron-pion scattering phase shifts.
We use the same approach as in the case of the proton in pion matter, replacing the proton by the deuteron, both treated as elementary particles.

To include the interaction of the deuteron with the dense pion system produced by the HIC, we use the deuteron-pion scattering phase shifts.
The Beth-Uhlenbeck formula is applied to the deuteron-pion channel and gives continuum correlations which may contribute to the observed deuteron yields.
A phase-shift analysis of pion-deuteron scattering has been performed in Ref.~\cite{Arvieux:1980pc,Rinat:1982ry,Garcilazo:1989bx}, 
calculations for the total and integrated elastic cross sections are given in \cite{Woloshyn:1976ca,Hiroshige:1984sm},
for a review of pion-deuteron scattering see \cite{Thomas:1979xu}. The pion-deuteron scattering amplitude is presented by Argand plots.
Of special interest is the p-wave amplitude calculated in different approximations \cite{Woloshyn:1976ca,Thomas:1979xu}.
The amplitude $f^2_{11}$ for the partial-wave $f^J_{LL'}$ with a total momentum $J=2$ and channel angular momentum $L, L'=1$ is compared to experimental data in Ref. \cite{Arvieux:1980pc}.

\begin{figure}[htb]
    \begin{center}
    \includegraphics[width = 0.48\textwidth]{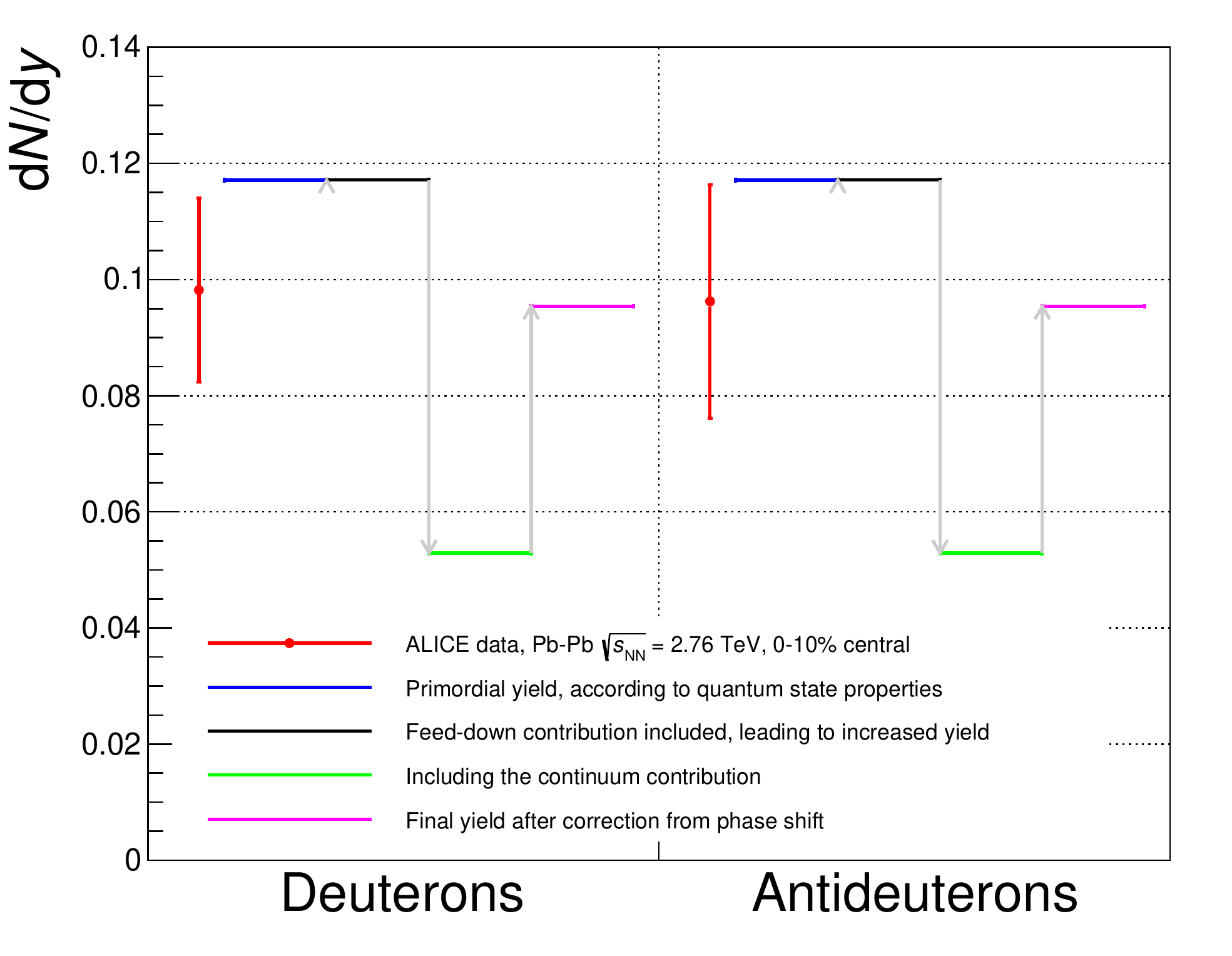}
    \end{center}
    \caption{Comparison between experimental data from Pb-Pb collisions at $\sqrt{s_\mathrm{NN}} = 2.76$~TeV for deuterons and antideuterons with model values. The red points indicate the production yield d$N$/d$y$ and the vertical lines attached to them the quadratic sum of the statistical and systematic uncertainties. The blue horizontal lines indicate the primordial yields, the black lines the yields corrected for feed-down contributions, the green lines include the contribution from the continuum and the magenta lines the yields after correcting the resonance contribution by the phase-shift analysis through the Beth-Uhlenbeck approach.}
 \label{fig:3}
\end{figure}

The partial-wave pion-deuteron scattering amplitudes $f^J_{LL'}$ are related to the strong phases $\delta^J_{LL'}$ and the $S$-matrix according to 
\begin{equation}
S=1+2 i k f^J_{LL'}=e^{2 i \delta^J_{LL'}}
\end{equation}
where $k$ is the momentum in the c.m. system. 
For exploratory calculations, we consider the amplitude $f^2_{11}$ which gives the largest contribution.
The following fit has been performed to the Argand plots of Refs. \cite{Woloshyn:1976ca,Arvieux:1980pc,Thomas:1979xu} to the phase shifts here, $E$ in MeV:
\begin{equation}
\tan \delta^2_{11}(E)=E^{3/2}\frac{13.5}{180^2-E^2}.
\end{equation}

The contribution of the pion-nucleon channel to the density is given by the intrinsic partition function, see Eq.~(\ref{sol_nucleon})
\begin{equation}
\label{sol_deuteron}
z_{d,\pi, L=1}=\frac{1}{\pi T_{\rm fr}} \int_0^\infty dE e^{-E/T_{\rm fr}}  \delta^2_{11}(E) = 0.312 
\end{equation}
(note that the phase shift is in radians).
With the degeneracy $g_\Delta=16$, the shift of the continuum by the pion mass, and the center-of-mass motion (for the nucleon-$\Delta$ resonance we take the value 2195.6 MeV) we obtain the contribution of the pion-deuteron channel to the density

\begin{eqnarray}
\label{deuterontotal}
&&\!\!\!\!\!\!\!\!\frac{dN_d^{\rm total}}{dy} = 0.0529 \left[1+ 4 \left(\frac{2195.6}{1876}\right)^{3/2}e^{-m_\pi/T_{\rm fr}}z_{d,\pi, L=1}\right]\nonumber \\ &&=0.0954
\end{eqnarray}
This value is near to the experimental result, see Fig. \ref{fig:3}. However, there are large uncertainties. 
Further pion-deuteron channels may be included, in particular $f^1_{11}$ and $f^3_{22}$ which will increase the deuteron-like density at freeze-out. Nevertheless, the contributions of the pion-deuteron scattering states may decay also to two nucleons as final states, considering the evolution of the fireball after freeze-out. For instance,
the inelasticities of pion-deuteron collisions are also given in the Argand diagrams. They will not influence the chemical equilibrium as long as detailed balance holds. Thermodynamic equilibrium is not realized after freeze-out, because reactive collisions become rare. If they happen, a branching ratio of about 0.25 holds for elastic scattering~\cite{Norem:1971nu}, so 0.75 are inelastic break up reactions. This way, pion-deuteron collisions determine the feed down of deuteron contributions after freeze-out, which decay to nucleons. Nevertheless, this can be largely neglected as contribution to the nucleon yield since the deuteron yield at LHC is by the factor of about 330 suppressed compared to the nucleon yield~\cite{Braun-Munzinger:2018hat}. The same argument can be used to explain that nucleon-nucleon (nn, pp, np) correlations are not essential for the proton yield, in contrast to the pion-proton correlations.

\section{Conclusions}
\label{Conclusions}

It is obvious that a simple statistical model (HRG) neglecting interaction effects cannot describe the composition of the hot and dense system produced by HIC experiments adequately. However, this simple model with 2 or 3 parameters does a rather good job in describing the yields of most particles. Aiming at precision requires clearly corrections to the simple model.
After the solution of the proton puzzle where the HRG overestimates the final proton yield~\cite{Andronic:2018qqt}, but the correct calculation of the scattering phase shifts gives the measured value, it was of interest to treat also other particles observed in the experiments. In the case of deuterons, the simple statistical model overestimates the deuteron yield seen in the experiment. The effect of pion-deuteron interaction in analogy to the HRG model would enhance the calculated deuteron yield because of feed-down from resonances like a $\Delta - N$ correlation, so that the discrepancy becomes larger. In the present work, before discussing the influence of the pion medium, we first explained that the deuteron yield is significantly reduced if the continuum correlations are included in accordance with the second virial coefficient. We conclude that this is the main mechanism to reduce the deuteron yield.

Similar effects are also expected for the other nuclei like triton or $^4$He. They are stronger bound, but the binding energy is also small compared to the temperature so that excited states and continuum correlations are relevant. 

The most extreme example, displayed also in Fig.~\ref{fig:pbm} and discussed intensively in~\cite{Braun-Munzinger:2018hat,Donigus:2020fon} is the hypertriton ${}^{3}_\Lambda$H, a bound state of a proton, a neutron and a $\Lambda$ hyperon. The object can be imagined as a deuteron core surrounded by the $\Lambda$ in form of a ultra-halo nucleus (size of about 10.8 fm~\cite{Braun-Munzinger:2018hat,Hildenbrand:2019sgp}). The hypertriton decays weakly and it's lifetime is expected to be close to the one of the free $\Lambda$ hyperon, since the $\Lambda$ separation energy from the deuteron core is only about 130 keV and the probability to find the $\Lambda$ far away from the core is high~\cite{Gal:2016boi}.
It would be outstanding to apply a phase shift correction as discussed here for the deuteron also to the hypertriton. Unfortunately is the data situation of scattering experiments for pions on the hypertriton even more scarse as for hyperons itself. Regrettably, the impulse approximation approach also discussed here did not lead to an acceptable result for the phase shift correction and therefore a different approach need to be found. 
It is also interesting to say that the correction needed would be in the opposite direction as the one for the deuteron and the ${}^{3}$He as visible from Fig.~\ref{fig:pbm}. 

It should be mentioned that the concept of a deuteron as a weakly bound two-nucleon system in a hot environment looks strange, see \cite{Braun-Munzinger:1994zkz,cern_courier,Mrowczynski:2016xqm,Oliinychenko:2018ugs,Mrowczynski:2020ugu,Vovchenko:2019aoz,Bugaev:2020sgz,Gallmeister:2020fiv,Neidig:2021bal,Oliinychenko:2020znl,Sun:2021dlz,Rais:2022gfg,Sun:2022xjr}. The rates of collisions to destroy or to form a nuclear bound state are large, 
but not of relevance in thermodynamic equilibrium  because of detailed balance. We have a large contribution from continuum correlations, as described by the two-nucleon spectral function. These correlations in the deuteron channel are considered as  precursors of the deuterons observed as the final deuteron yields. A similar concept is used in the coalescence model~\cite{butler_pearson61,butler_pearson63,Sato:1981ez,Remler:1981du,Gyulassy:1982pe,Kapusta:1984ij,Csernai:1986qf,Kolybasov:1989rfy,Mrowczynski:1989jd,Koch:1990psq,Dover:1991zn,Danielewicz:1991dh,Mrowczynski:1992gc,Hirenzaki:1992gx,Leupold:1993ms,Nagle:1994wj,Llope:1995zz,Nagle:1996vp,Mattiello:1995xg,Bleicher:1995dw,Nagle:1996vp,Scheibl:1998tk,Steinheimer:2012tb,Sun:2015jta,Sun:2015ulc,Feckova:2015qza,Feckova:2016kjx,Sun:2017ooe,Botvina:2017yqz,Tomasik:2017mbf,Blum:2017qnn,Bellini:2018epz,Sombun:2018yqh,Zhao:2018lyf,Sun:2018mqq,Kachelriess:2019taq,Bleicher:2019oog,Blum:2019suo,Donigus:2020fon,Sun:2020uoj,Gaebel:2020wid,Gaitanos:2021cfh,Hillmann:2021zgj,Reichert:2021ljd,Sochorova:2021lal,Glassel:2021rod,Zhao:2021dka}. Baryon pre-clusters are also discussed in~\cite{Shuryak:2018lgd,Shuryak:2019ikv,Shuryak:2020yrs} in a slightly different approach.

As mentioned above, a more detailed description of the expansion process should consider also the fate of these correlations in the 
expanding system, for instance the evolution of the two-nucleon spectral function with time when the thermodynamic parameters of the high density environment are changing. The hydrodynamic stage of the evolution of the fireball is based on a description which assumes local thermodynamic equilibrium where the treatment of correlations (bound states, resonances, continuum correlations) is possible. We assume that this is an appropriate approximation until chemical freeze-out where the primordial yields are formed. To go beyond the hydrodynamic stage, the kinetic stage gives the appropriate description where the relevant observables are the distribution functions in momentum space. Observables such as the transverse momentum spectra for protons and deuterons are obtained from transport model calculations. A hydrodynamical blast-wave model may reproduce some signatures of these spectra but is not subject of the present work. With respect to the yields, we assume that the final composition is obtained from the primordial composition after taking into account feed-down processes occurring in the kinetic stage of evolution. Further work is necessary to improve this approximation and to compare with transport model calculations. 

The description of the hadronic phase is a strength of the transport models, such as the quantum molecular dynamics. In fact, they are kind of afterburners on the hydrodynamic phase, namely they can incorporate things as absorption or annihilation of particles with great success, see for instance~\cite{Stock:2018xaj} and references therein.  
The question of freeze-out probes in heavy-ion collisions has been discussed also in context with strange hadron resonances \cite{Markert:2002rw}. 
Suppression of strange particle resonance production at LHC energies has been discussed recently as result of the hadronic phase while expansion \cite{ALICE:2018ewo,Motornenko:2019jha,ALICE:2020mkb,ALICE:2021ovi,Knospe:2021jgt,Oliinychenko:2021enj,Oliinychenko:2022xbh}. 
Short-lived resonances, scattering and rescattering processes may help to achieve a better description of the expanding fireball after freeze-out.
Feed-down concepts based on reaction networks are only approximations, a systematic treatment should be obtained from a non-equilibrium statistical operator approach.

The correct description of correlations in the relevant statistical operator, given up to freeze-out by the local thermodynamic equilibrium, is a prerequisite to describe the non-equilibrium evolution of the fireball produced by heavy-ion collisions. 
Signatures of resonances, excited states, continuum correlations are also seen in the observed, final distribution of nucleons and nuclei.

\section*{Acknowledgements}
We thank Johanna Stachel and Peter Braun-Munzinger for the hospitality and many useful discussions. 
We are also grateful to  Volodymyr Vovchenko for helpful correspondance and for reading and commenting the paper draft. We also appreciate discussions with Volodymyr Vovchenko on the feed-down contribution of excited light nuclei. We further thank Anton Andronic and Pok Man Lo for the exchange about the S-matrix approach and the corresponding fit result. 

This research was supported in part by the ExtreMe Matter Institute EMMI at the GSI Helmholtzzentrum f\"{u}r Schwerionenforschung, Darmstadt, Germany. B.D. acknowledges the support from Bundesministerium f\"{u}r Bildung und Forschung through ErUM-FSP T01 (F\"{o}rderkennzeichen 05P21RFCA1). D.B. was supported by NCN under grant No. 2019/33/B/ST9/03059.

\appendix
\section{Deuteron-like correlations}
\label{lab:appendix}

The Beth-Uhlenbeck formula for the second virial coefficient (\ref{BU}) has to be generalized to implement the quasiparticle picture. We show how deuteron-like correlations are extracted from this approach. We consider the nucleon system at temperature $T$ and chemical potentials $\mu_n,\mu_p$.

The quantum statistical approach to correlations in nuclear matter~\cite{Schroder:2017dww} considers correlation functions and its Fourier transform, the spectral function $S_\tau(1,\omega;T,\mu_n,\mu_p)$.
 The single-nucleon quantum state $|1\rangle$
can be chosen as $1 = \{{\bf p}_1, \sigma_1,\tau_1\}$ which denotes wave number, spin, and isospin, respectively.
A rigorous expression for the nuclear matter equation of state is found provided that the spectral function is known,
\begin{equation}
\label{EOS}
  n^{\rm tot}_\tau(T,\mu_n,\mu_p)=\frac{1}{\Omega}\sum_{p_1,\sigma_1} \int \frac{d \omega}{2 \pi} \frac{1}{e^{(\omega-\mu_\tau)/T}+1}
  S_\tau(1,\omega)
\end{equation}
($ \Omega$ is the system volume, $\tau = \{n,p\}$; we take $k_B=1$).
The spectral function $S_\tau(1,\omega;T,\mu_n,\mu_p)$
is related to the self-energy $\Sigma(1,z)$ for which a systematic Green functions approach is possible using diagram techniques:
\begin{equation}
\label{spectral}
 S_\tau(1,\omega) = \frac{2 {\rm Im}\Sigma(1,\omega-i0)}{
(\omega - E(1)- {\rm Re} \Sigma(1,\omega))^2 +
({\rm Im}\Sigma(1,\omega-i0))^2 }\,;
\end{equation}
$E(1)=p_1^2/2m_N$.

For the self-energy a cluster decomposition is possible \cite{Ropke:2020hbm}.
As shown in Sec. \ref{Partial},  the total density is decomposed
into partial contributions from different channels (\ref{ntotal}).
In particular, the two-nucleon contribution $A=2$ is given by the Beth-Uhlenbeck formula for the second virial coefficient (\ref{BU}).
However, the structure of the spectral function (\ref{spectral}) leads to the quasiparticle picture.
For small $ {\rm Im}\Sigma(1,\omega-i0)$, the pole appears at the quasiparticle energy solving $E^{\rm qu}(1)=E(1)+ {\rm Re} \Sigma(1,E^{\rm qu}(1))$.
As consequence, the single-particle contribution ($A=1$) reads
\begin{equation}
n^{\rm part}_{1,\tau}=2 \int \frac{d^3 P}{(2 \pi)^3} \frac{1}{e^{(E^{\rm qu}(1)-\mu_\tau)/T}+1}.
\end{equation}

A well-known example is the Hartree-Fock approximation or, more general, the mean-field approximation, which is a standard approach to nuclear systems.
However, part of the interaction which is described within the second virial coefficient is already used to introduce the quasiparticle picture.
A systematic approach has been given in~\cite{Schmidt:1990oyr}, and a generalized Beth-Uhlenbeck formula has been derived:
\begin{eqnarray}
\label{gBU}
&&z_{np} ^{\rm part}(P;T,\mu_n,\mu_p)\nonumber \\ && =
3e^{-\frac{E_{np}^{\rm cont}( P)}{T}} \left[\left(e^{B_d( P)/T}-1\right)\right]\Theta[B_d( P)]\nonumber \\ &&
+\frac{1}{\pi T}\int dE e^{-E/T}\left\{\delta_{np}(E)-\frac{1}{2}\sin[2\delta_{np}(E)]\right\}.
\end{eqnarray}
The edge of continuum $E_{np}^{\rm cont}( P)=2 E^{\rm qu}(P/2)- P^2/(4 m_N)$ is different from zero if quasiparticle energies are considered. $B_d( P)$ is the deuteron binding energy  which in general is medium modified and thus depending on $P$ as well as temperature $T$ and chemical potentials of components, the same holds also for the scattering phase shifts $\delta_{np}(E)$. $E$ is the energy in the c.m. system.
$\Theta[x]$ is the step function. The last term $\sin[2\delta_{np}(E)]$ is necessary to avoid double counting, because part of the interaction is already taken into account in the Hartree-Fock quasiparticle shifts of the nucleons~\cite{Schmidt:1990oyr}. It needs a special discussion of the spectral function as function of the frequency to decide which part is included in the quasiparticle contribution. The remaining part describes correlations in the system. 

For strong interactions, the Hartree-Fock approximation for the quasiparticle energies has to be improved. 
For instance, nucleon-nucleon interaction at short distances is strongly repulsive~\cite{Typel:2016srf,Burrello:2022tjw}. 
This dominates the properties at high energies and leads to the concept of excluded volume, 
i.e. free motion of the particles at apparent higher density or shifted chemical potential.
The comparison of the hadron gas with a hard-core nucleon model has been performed 
in ~\cite{Vovchenko:2017drx}. Different approaches including lattice QCD results
near $T=150$ MeV infer a value of about $r_c=0.3$ fm for the hard core radius.

The deuteron appears as bound state in the isoscalar $^3S_1$ channel. 
There is a small admixture (4 \%) from the $^3D_1$ channel because of the tensor forces which will be neglected here. If we use the quasiparticle description of nucleons with hard-core interaction potential, the phase shift for the $S$ state (angular momentum 0) $\delta^{\rm hc}_0(E^{\rm lab})=-2 r_c (m_N E^{\rm lab}/2 )^{1/2}$
is included in the quasiparticle contribution of the spectral function. 
For the density contribution of the deuteron channel remains the virial coefficient 
\begin{eqnarray}
&& b^{\rm qu}_d(T)=\frac{3}{2^{1/2}}\left[e^{B_d/T}-1 \right. \\ && \left.
+\frac{1}{2 \pi T}
\int_0^\infty dE^{\rm lab} e^{-E^{\rm lab}/2T} \left(\delta_{^3S_1}(E^{\rm lab})-\delta^{\rm hc}_0(E^{\rm lab}) \right)\right].\nonumber 
\end{eqnarray}
Using the SAID nucleon-nucleon phase shifts~\cite{Workman:2016ysf,said:ws}, the value $b^{\rm qu}_d(T_{\rm fr})=0.9713$ is obtained.


\nocite{*}

\bibliography{refs}

\end{document}